\documentclass[prb,aps,amssym,nofootinbib,floatfix,twocolumn,notitlepage, eqsecnum]{revtex4-1} 

\usepackage{placeins}
 \usepackage{xcolor}
\usepackage{amsmath,amssymb,bbold,bm}
\usepackage{graphicx}
\usepackage{caption}
\usepackage{subcaption}
\usepackage{cancel}
\usepackage{comment}

\newcommand{\be}{\begin{equation}}
\newcommand{\ben}{\begin{equation*}}
\newcommand{\ee}{\end{equation}}
\newcommand{\een}{\end{equation*}}
\newcommand{\bs}{\begin{split}}
\newcommand{\es}{\end{split}}
\newcommand{\bmx}{\begin{array}}
\newcommand{\emx}{\end{array}}
\newcommand{\bea}{\begin{eqnarray}}
\newcommand{\bean}{\begin{eqnarray*}}
\newcommand{\eea}{\end{eqnarray}}
\newcommand{\eean}{\end{eqnarray*}}
\newcommand{\dg}{^{\dagger}}
\newcommand{\dn}{^{\vphantom{\dagger}}}

\newcommand{\lr}{\leftrightarrow}

\newcommand{\ua}{\uparrow}
\newcommand{\da}{\downarrow}
\newcommand{\bb}[1]{\mathbb{#1}}

\newcommand{\andd}{\qquad\text{and}\qquad}

\newcommand{\eps}{\epsilon}

\newcommand{\sgn}[1]{{\rm sign}{#1}}
\newcommand{\pref}[1]{(\ref{#1})}

\newcommand{\abs}[1]{\left\vert #1 \right\vert}

\newcommand{\braket}[1]{\left\langle #1\right\rangle}

\newcommand{\mat}[1]{\left(\bmx{cc}#1\emx\right)}
\newcommand{\matn}[1]{\bmx{cc}#1\emx}
\newcommand{\matl}[1]{\bmx{ll}#1\emx}

\newcommand{\bw}[1]{\begin{widetext}}
\newcommand{\ew}[1]{\end{widetext}}

\begin{document}
\title{Detecting a quantum critical point in topological SN junctions}
\author{Yashar Komijani$^{1,2}$}\email{komijani@phas.ubc.ca}
\author{Ian Affleck$^1$}
\affiliation{ $^1$Department of Physics and Astronomy and $^2$Quantum Materials Institute, University of British 
Columbia, Vancouver, B.C., Canada, V6T 1Z1}
\date{\today}
\begin{abstract}
A spin-orbit coupled quantum wire, with one end proximate to an s-wave superconductor, can become a topological superconductor, with 
 a Majorana mode localized at each end of the superconducting region. It was recently shown that coupling 
one end of such a topological superconductor to {\it two} normal channels of interacting electrons leads to 
a novel type of frustration and a quantum critical point when both channels couple with equal strength. We propose 
an experimental method to access this critical point in a {\it single} quantum wire and show  its resilience to disorder.

\end{abstract}
\maketitle
\section{Introduction}
There has recently been great interest in topological superconductor (TSC) quantum wires, which are predicted 
to host localized Majorana modes (MM's) at their ends.\,\cite{Kitaev01,Oreg10,Lutchyn10} Apart from being intrinsically fascinating, these exotic objects have 
potential applications as topologically protected qubits.\,\cite{Alicea11,Alicea12} 
Experiments on quantum wires with strong spin-orbit interaction (SOI), proximate 
to a superconductor and in a magnetic field, have shown some evidence for the MM, predicted to exist 
in the quantum wire where it extends past the edge of the superconductor.\,\cite{Mourik12,Das12,Deng12} Topological SN junctions are predicted 
to exhibit perfect Andreev scattering at low energies and a corresponding zero bias peak (ZBP) has been detected. 
However, other explanations of this anomaly have been proposed and the long-sought MM remains elusive.\,\cite{Pikulin12}

An SN junction between a topological superconductor and {\it two} normal channels was studied theoretically in Ref.\,\onlinecite{AG}.
This could correspond to a Y-junction between a TSC and two normal quantum wires.  
It was shown that, due to electron-electron interactions,
 a novel type of quantum frustration occurs at low energies when the two channels are coupled to the 
MM with near-equal strength. Perfect Andreev reflection occurs in the more strongly coupled channel 
and perfect normal reflection in the more weakly coupled channel. This implies that the MM Y-junction 
acts as a very sensitive switch, a property of possible importance if a network of such junctions 
is assembled for quantum information processing. When the two channels couple with precisely equal 
strength, the frustration leads to a novel quantum critical point (QCP). An alternative  realization of 
this 2-channel topological SN junction could be possible in a single quantum wire extending past 
the end of a superconductor, as in previous experiments, {\it provided that both spin channels couple with equal strength to the MM}.
Due to the magnetic field, it is generally 
expected that only one channel couples to the MM, as discussed below.

We show here that it is possible to 
couple both spin channels to the MM if the magnetic field changes abruptly over a length scale $L_{\rm SN}$ near the SN junction much smaller than the nanowire length $L_{\rm NW}$.  By tuning the magnetic field,  
the coupling of the MM to both channels can be made equal and the quantum critical point studied 
by transport measurements. In Sec. II we  discuss our experimental proposal and demonstrate its 
viability by numerical simulations of a clean non-interacting quantum wire. For non-interacting two-channel quantum wires we have recently shown~\cite{KA} that the conductance is $2e^2/h$ independent of the coupling strength of each channel to the MM and independent of disorder near the junction.
In Sec. III we observe that deviations from this universal value of the conductance at low energies 
as the magnetic field is varied provide 
a signature of the interactions-induced quantum critical point. A number of Appendices provide additional details and discuss numerical methods and further experimental considerations.

 \begin{figure}[h!]
\includegraphics*[width=1.\linewidth]{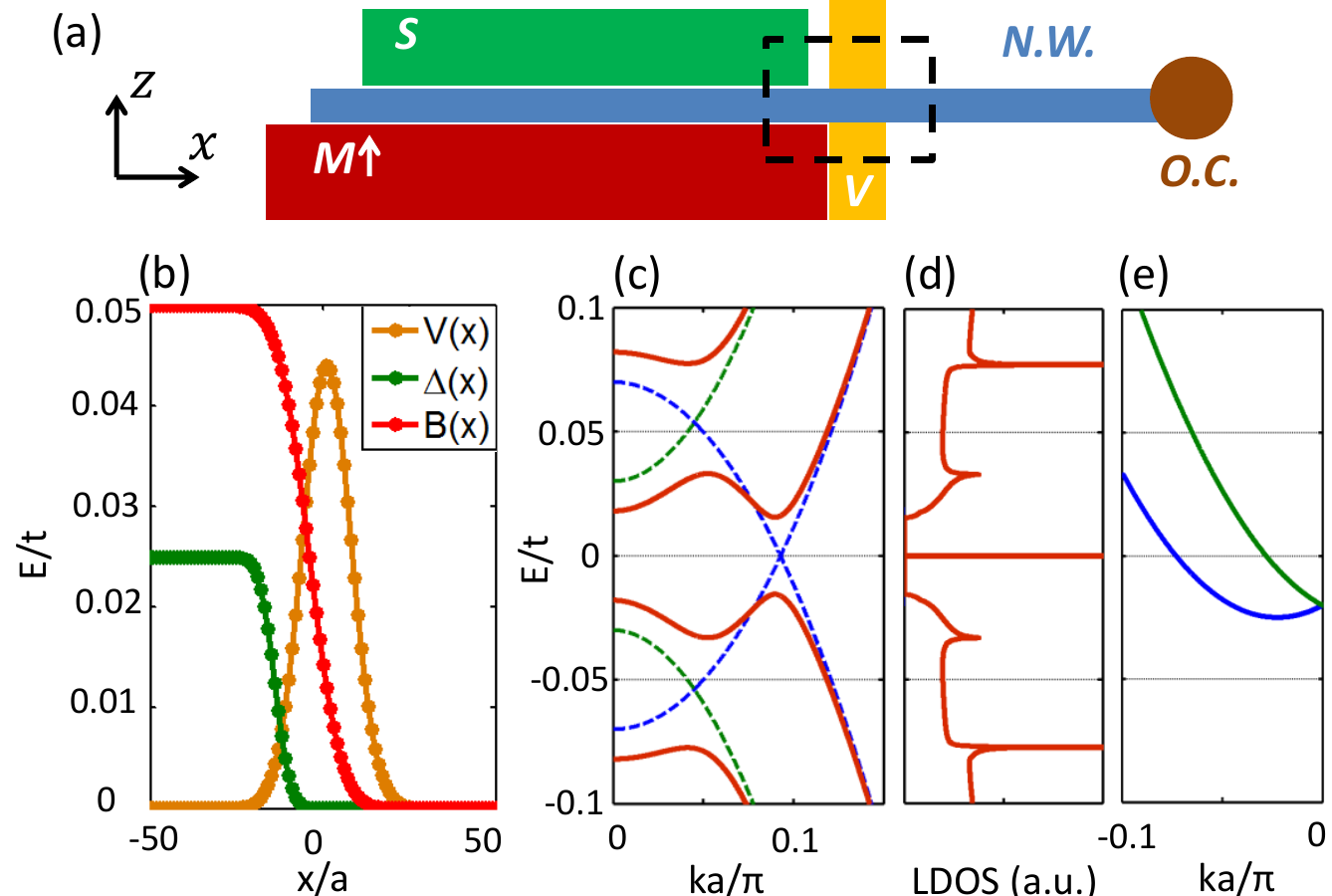}
\caption{
\raggedright\small
(color online) (a) Schematic of the proposed setup. A nanowire (N.W.) is contacted by an Ohmic contact (O.C.). A superconducting layer on top provides the proximity pairing and a magnetic layer underneath causes a phase transition into a TSC regime which ensures presence of a MM in the nanowire above the pinch off gate ($V$). (b) Parameters of the corresponding tight-binding model, within the dashed box of (a) normalized to the half-bandwidth, $t$. The chemical potential in the tight-binding model is 
$\mu_{{tb}} = -1.98t$ so that a good approximation to a continuum model is obtained with $\mu =0.02t$. 
 We assume an external magnetic field $B_{ext}$ in addition to the field produced by the magnetic layer shown here, with 
both fields parallel. Moreover, a constant SOI energy $E_{SO}/t=m\alpha^2/2t\hbar^2\sim 0.05$ is assumed. Parameters of the model vary within a length scale $L_{\rm SN}$ much shorter than the length of the wire $L_{\rm NW}$ but slowly 
compared to the lattice constant, $a$.  (c) Dispersion curves on the superconducting side. The red curve and the blue/green curves show the dispersions w/o including the proximity coupling. (d) Local density of states at the end of TSC in the limit of infinite $V$, showing a MM at $E=0$. (e) Dispersion in the normal side of the nanowire shows the two channels at the Fermi energy.} \label{fig:setup}
\end{figure}
\section{The theoretical model and the Experimental proposal}

Keeping only 2 spin-split channels, the Hamiltonian for our proposed device, which is sketched in Fig.\,\ref{fig:setup}, can be written
\bw

\be
H=\int {dx}\Big\{\psi\dg (x)\Big[{p^2\over 2m}-\mu+V(x)+\alpha p \sigma^y+B(x)\sigma^z\Big]\psi (x)
+\Big[\Delta (x)\psi_\uparrow (x)\psi_\downarrow (x)+h.c.\Big]\Big\}+H_{int}.\label{Hq}
\ee

\ew

Here $p=-i\hbar\partial_x$ and a sum over spin indices is implicit in the $\psi^\dagger \psi$ term. We work in units 
where $g\mu_B=2$, so $B(x)$ is the Zeeman energy. 
The potential barrier, $V(x)$, controls the transparency of the SN junction. 
The Rashba SOI corresponds to an electric field in the $z$-direction, between wire and the substrate. 
The actual directions of the magnetic field and spin-orbit vector, $\vec \alpha$ are unimportant provided that 
they are orthogonal to each other; in that case the Hamiltonian can always be mapped to the form of Eq.\,(\ref{Hq}) by a 
unitary transformation.\,\cite{sob}
The magnetic field $B(x)$ and the proximity induced pairing strength of the quantum wire $\Delta (x)$, are assumed to vary rapidly,  over a distance $L_{\rm SN}$, near $x=0$.
 We must choose $\mu$, $B(x\ll -L_{\rm SN}/2)\equiv B_S$ and $\Delta (x\ll -L_{\rm SN}/2)\equiv \Delta$ to 
obey $B_S>\sqrt{\Delta^2+\mu^2}$, so that 
the superconducting portion of the wire is in the topological phase with a MM located near $x=0$. This is indicated by the local density 
of states (LDOS) just to the left of the barrier, for the case of infinite barrier, $V\to \infty$,  sketched in Fig.\,\ref{fig:setup}(d).  
The energy bands in the normal region are given by
\be E_{\pm}(p)= {p^2\over 2m}-\mu\pm \sqrt{(\alpha p)^2+B^2} \label{EB}\ee
and the corresponding 2-component wave-functions are $e^{ipx}\chi_{p,\pm}$ where the two component spinors $\chi_{p,\pm}$ are given explicitly in Appendix \ref{sec:BdG}. Thus, if $B$ is uniform, only one band is occupied in the normal region. 
However, if $B(x\gg L_{\rm SN}/2)\equiv B<\mu$, both bands are partially occupied and it is possible to access the QCP.
The needed abruptly changing magnetic field might be achieved by proximity to a ferromagnetic layer terminating at $x=0$, with 
spins oriented perpendicular to the quantum wire. Another possibility might be to cloak the quantum wire in a 
thin superconducting layer which might act as a nano-solenoid and could trap some flux during cool down. 

We calculate the sum of the current flowing in both channels in the normal region, far to the right of the SN junction, 
at zero temperature, 
in linear response to a voltage applied at $x\gg L_{\rm SN}/2$ to either channel. The corresponding linear conductances at $T=0$, 
ignoring interactions, are \cite{BTK}
\be G_i=(e^2/h)\left[ 1-\left(  {\bb r}^{ee} {\bb r}^{\dagger ee} \right)_{ii}
+\left(  {\bb r}^{he} {\bb r}^{\dagger he}\right)_{ii} \right] \label{Gi}.\ee
Here $ {\bb r}^{ee}_{ji}$ and $ {\bb r}^{he}_{ji}$ are $2\times 2$ matrices, in the space of the two channels. 
They give the amplitude for an incoming electron of zero energy in channel $i$ to be scattered as an electron or hole 
in channel $j$, multiplied by a factor of $\sqrt{v_i/v_j}$ where $v_i$ are the Fermi velocities for each channel   
(Appendix \ref{sec:BdG}).
\subsection{Feasibility of tuning to the equal-coupling point}

In the limit of weak tunnelling between superconducting and normal side, and therefore weak coupling of the MM to the normal wire, we find 
that both channels couple with equal strength to the MM, when the magnetic field is zero on the normal side. This limiting result can be understood by the fact that in this regime, the coupling is simply the overlap between wavefunctions of the MM and the two gapless channels with proper boundary conditions, each obtained separately in absence of the other. In absence of the magnetic field the two channels in the normal region are eigenstates of $\sigma^y$, meaning that they can be written as
\be
\vec \Psi_1(x)=\mat{1\\i}f(x)\ \ 
\vec \Psi_2(x)=\mat{1\\-i}f(x).
\ee
On the other hand since the BdG Hamiltonian is real (Appendix \ref{sec:BdG}) the MM wave-function can be chosen to be real. Denoting the electron part of the MM wavefunction by the real vector $\vec w^e(x)$, it follows that 
\bea
&& \bigg| \int dx\vec w^{e*}(x)\cdot \vec \Psi_1(x) \bigg|^2=\bigg| \int dx \vec w^{e*}(x)\cdot \vec \Psi_2(x)\bigg|^2 \nonumber \\
&&=\left(\int dx w^e_1(x)f(x)\right)^2+ \left(\int dx w^e_2(x)f(x)\right)^2
\eea
and thus $\abs{t_1}^2=\abs{t_2}^2$. It can be shown~\cite{KA} that this is equivalent to having equal conductances for the two channels $G_1=G_2$ in the non-interacting case. Remarkably, this proof only assumes that the SOI is perpendicular to the magnetic field and has the same direction (not magnitude)
 on both S and N sides. 
Our numerical results indicate $G_1=G_2$ also for a high smoothly varying barrier, $V(x)$, leading to a weak coupling to the MM. 
However, the $\abs{t_1}^2=\abs{t_2}^2$ condition is destroyed by a stray field on the normal side or 
by increasing the coupling to the MM, causing the lower subband [blue in Fig.\,\ref{fig:setup}(e)] to couple more strongly to the polarized Majorana. Fortunately, this asymmetry can be compensated by applying a small external field of the opposite sign as confirmed by our numerical results.
\subsection{Numerical analysis}\label{sec:num}
Different numerical methods have been used to calculate the conductance of topological SN junctions.\,\cite{Lin12,Prada12,Pientka12,Rainis13,Chevallier13} We use an exact diagonalization method augmented by recursive calculation of self-energies, to calculate the differential conductance of a tight-binding model whose parameters are depicted in Fig.\,\ref{fig:Gschem}(b). We have chosen the density and the rate of variation of the parameters to be $\ll 1/a$, 
where $a$ is the lattice spacing, so that a 
good approximation to the continuum model is obtained. The details of the method and the parameters used are described in Appendix \ref{sec:num}.

\begin{figure}[h!]
\includegraphics*[width=1\linewidth]{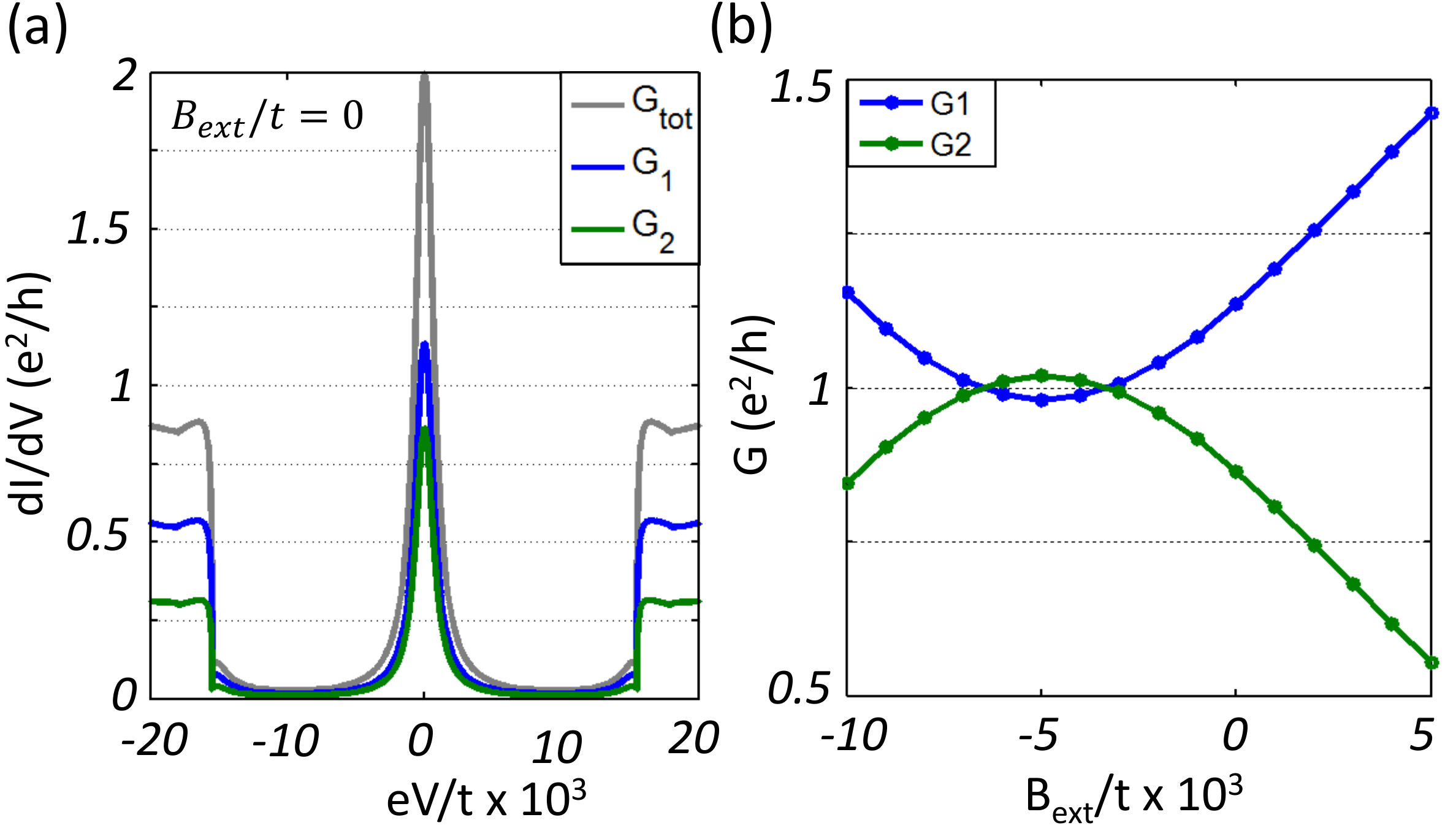}
\caption{\raggedright\small (color online) (a) The zero-temperature $(T=0)$ channel resolved differential conductance of the SN junction, computed for parameters in Fig.\,\ref{fig:setup}(b), exhibits a ZBA inside the gap.  The blue/green subbands correspond to incident electrons in blue/green subbands of Fig.\,\ref{fig:setup}(e). The total conductance (gray) peaks to $2e^2/h$ at $V=0$ and is independent of other parameters as expected. (b) $G_1$ (blue) and $G_2$ (green) can be tuned by $B_{ext}$, becoming equal 
 at $B_{ext}/t\approx -5\times 10^{-3}$.}
\label{fig:Gfree}
\end{figure}

\begin{figure}
\includegraphics[width=1.\linewidth]{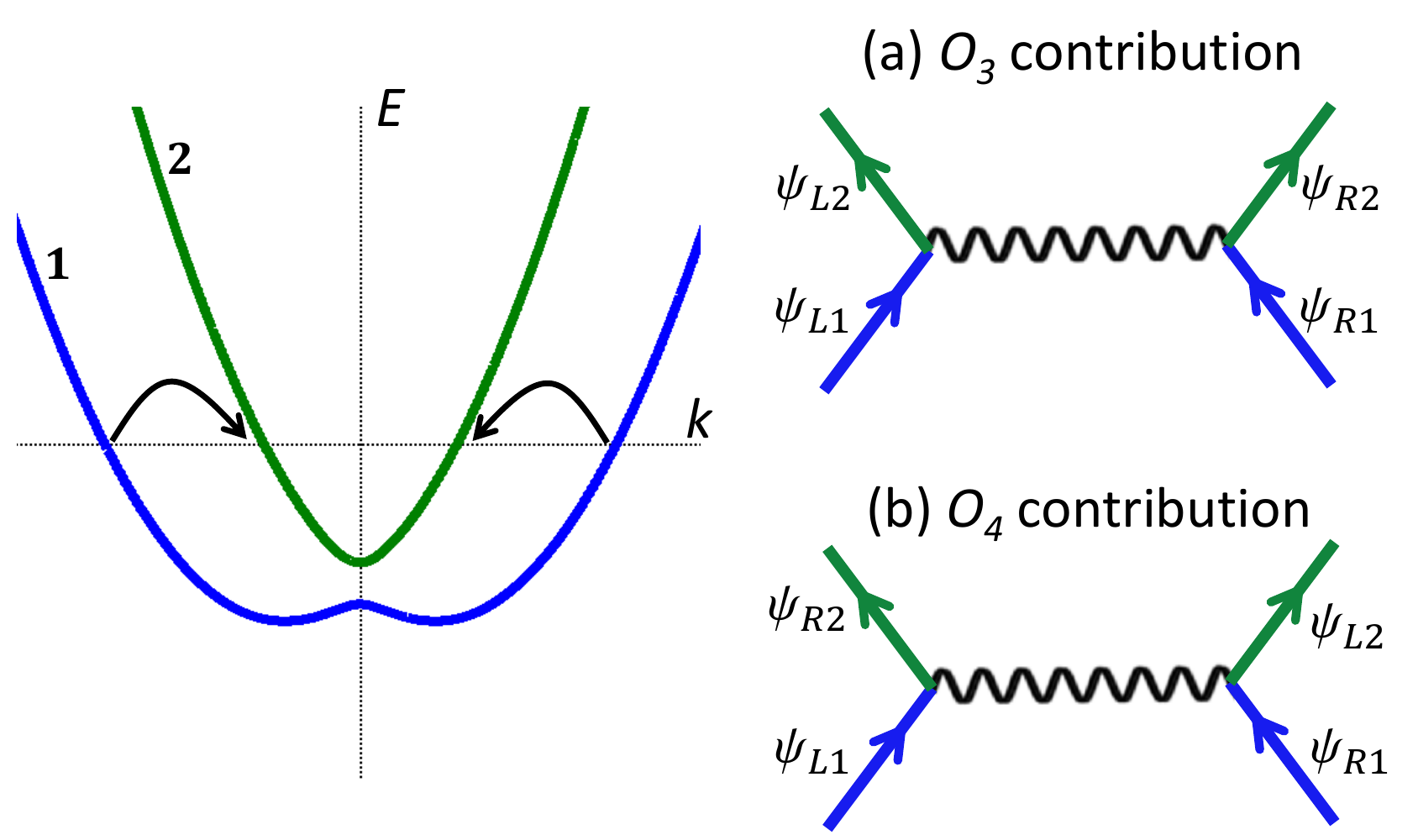}
\caption{
\raggedright\small (color online) The two momentum-conserving contributions to double channel-flip scattering processes in Eq.\,\pref{eqfH}. $O_3$ needs magnetic field and SOI to be present. $O_4$ exists without a magnetic field.\label{fig:Zprocess}
}
\end{figure}

The results of such a calculation for the parameters of Fig.\,1(b) are shown in Fig.\,{\ref{fig:Gfree}. Fig.\,2(a) shows the computed channel-resolved differential conductance as well the total conductance at zero temperature, which all exhibit a zero bias peak inside the gap. The total conductance becomes $2e^2/h$ at zero bias as expected from topological arguments.~\cite{KA} We have chosen $B(x)=B_M(x)+B_{ext}$ where 
the function $B_M(x)$ is a smoothed out step function representing the contribution of the ferromagnet and 
$B_{ext}$ is an adjustible parameter corresponding to an externally applied field. Note that the zero bias conductance are not the same when external magnetic field is zero. As $B_{ext}$ is varied, we see 
from Fig.\,\ref{fig:Gfree}(b) that points are passed through where $G_1=G_2$. According to Ref.\,\onlinecite{KA}, this is a sufficient condition for equal coupling.


\section{Interacting case and the access to the quantum critical point}\label{sec:int}
We now consider the effects of electron-electron interactions in the normal part of the wire, turning it into a 2-channel Luttinger liquid (LL) and show that the low-energy Hamiltonian of the normal side reduces to the form considered in Ref.\,\onlinecite{AG}. We start with Eq.\,(1) and the interaction
\be H_{int}=\int{dxdy V(x-y)\rho(x)\rho(y)}\label{Hint}
\ee
Denoting two Fermi momenta with $k_{Fj}$, the Fermi velocities with $v_{Fj}$ and the wavefunctions on the space of two modes with the short-hand notation $\chi_j\equiv\chi_{jk_{Fj}}$,
\bea
\Psi(x)&=&e^{ik_{F1}x}\psi_{R1}(x)\chi\dn_{1}+e^{-ik_{1}x}\psi_{L1}(x)\chi_{1}^*\\
&+&e^{ik_{F2}x}\psi_{R2}(x)\chi\dn_{2}+e^{-ik_{2}x}\psi_{L2}(x)\chi_{2}^*
\eea
where $\psi_{Rj}$ and $\psi_{Lj}$ are right/left movers of channel $j$. Inserting this into the Eq.\,\pref{Hq} for $x>0$ we arrive at the Hamiltonian density
\bw

\bea
\mathcal{H} &\approx &
iv_{F1}\Big(\psi\dg_{R1}\partial_x\psi\dn_{R1}-\psi\dg_{L1}\partial_x\psi\dn_{L1}\Big)
+iv_{F2}\Big(\psi\dg_{R2}\partial_x\psi\dn_{R2}-\psi\dg_{L2}\partial_x\psi\dn_{L2}\Big)
-2\tilde V(2k_1)\abs{O_1}^2\Big[\psi\dg_{L1}\psi\dn_{L1}\psi\dg_{R1}\psi\dn_{R1}\Big]
\nonumber\\
&
-&2\tilde V(2k_2)\abs{O_2}^2\Big[\psi\dg_{L2}\psi\dn_{L2}\psi\dg_{R2}\psi\dn_{R2}\Big]+\tilde V(0)\Big[\psi_{R1}\dg\psi\dn_{R1}+\psi\dg_{L1}\psi\dn_{L1}
+\psi_{R2}\dg\psi\dn_{R2}+\psi\dg_{L2}\psi\dn_{L2}\Big]^2\nonumber\\
&-&2\tilde V(2k_1)\abs{O_1}^2\Big[\psi\dg_{L1}\psi\dn_{L1}\psi\dg_{R1}\psi\dn_{R1}\Big]
+2\tilde V(k_2-k_1)\abs{O_3}^2\Big(\psi_{R1}\dg\psi\dn_{R2}+\psi\dg_{L2}\psi\dn_{L1}\Big)\Big(\psi_{R2}\dg\psi\dn_{R1}+\psi\dg_{L1}\psi\dn_{L2}\Big)\nonumber\\
&+&
2\tilde V(k_2+k_1)\abs{O_4}^2\Big(\psi\dg_{L1}\psi\dn_{R2}+\psi\dg_{L2}\psi\dn_{R1}\Big)\Big(\psi_{R2}\dg\psi\dn_{L1}+\psi\dg_{R1}\psi\dn_{L2}\Big)\label{eqfH}
\eea
\ew
where $\tilde{V}(k)$ is the Fourier transform of the interaction potential $V(x)$. The overlap parameters $O_n$ are given by
\bea
\matl{
O_1=\Sigma_{\sigma}\chi^2_{1\sigma} &\qquad
O_2=\Sigma_{\sigma}\chi^2_{2\sigma}\\
O_3=\Sigma_\sigma\chi\dn_{2\sigma}\chi_{1\sigma}^* &\qquad
O_4=\Sigma_\sigma\chi\dn_{2\sigma}\chi_{1\sigma}}\label{eq6}
\eea
Using bosonization
\be
\psi_{Lj/R j}(x)\propto e^{i\sqrt{\pi}[\varphi_j(x)\pm\theta_j(x)]},
\ee
with $\varphi$ and $\theta$ bosons satisfying
\be
[\varphi_i(x),\theta_j(y)]=-\frac{i}{2}\delta_{ij}\sgn(x-y)\label{eq48}
\ee
we arrive at
\bea
\mathcal{H}&=&\frac{1}{2}\sum_{j=1,2}u_j\Big[K\dn_j(\partial_x\varphi_j)^2+K_j^{-1}(\partial_x\theta_j)^2\Big]\nonumber\\
&+&U_{\varphi}\partial_x\varphi_1\partial_x\varphi_2+U_{\theta}\partial_x\theta_1\partial_x\theta_2\vphantom{\Big[}\nonumber\\
&+&\eta{Z}\cos 2\sqrt{\pi}(\varphi_1-\varphi_2).\vphantom{\Big[}\label{eqbos}
\eea
This is similar to the Hamiltonian of Ref.\,\onlinecite{AG} except the additional $U_{\varphi}$ (which was set to zero there) and the sine-Gordon $Z$-term. This term is produced here by double channel-flip scattering process with the fermionic representation of $\psi\dg_{R2}\psi\dg_{L2}\psi\dn_{R1}\psi\dn_{L1}+h.c.$ in the last two lines of Eq.\,\pref{eqfH} and $\eta$ is a cut-off dependent constant (see below). Other parameters are
\bea
u_jK_j&=&v_{Fj}+\frac{\tilde{V}(2k_j)\abs{O_j}^2}{\pi}, \\ u_jK_j^{-1}&=&v_{Fj}+\frac{2\tilde V(0)-\tilde{V}(2k_{F1})\abs{O_j}^2}{\pi}
\\
&&\hspace{-1.5cm}Z=\frac{-\tilde{V}(k_{F1}-k_{F2})\abs{O_3}^2+\tilde{V}(k_{F1}+k_{F2})\abs{O_4}^2}{\pi}\nonumber
\eea
and
\bea
U_{\theta}&=&\frac{2\tilde{V}(0)-\tilde{V}(k_{F1}-k_{F2})\abs{O_3}^2-\tilde{V}(k_{F1}+k_{F2})\abs{O_4}^2}{\pi}\nonumber\\
U_{\varphi}&=&\frac{-\tilde{V}(k_{F1}-k_{F2})\abs{O_3}^2+\tilde{V}(k_{F1}+k_{F2})\abs{O_4}^2}{\pi}
\eea
Ref.\,\onlinecite{AG} diagonalizes this Hamiltonian by introducing  new  $\sigma,\rho$ bosons which are related to the original bosons by
\be \left(\begin{array}{c}r\phi_1\\r^{-1}\phi_2\end{array}\right)=O(\beta )\left(\begin{array}{c}\phi_\sigma\\ \phi_\rho\end{array}\right),\ \ 
 \left(\begin{array}{c}r^{-1}\theta_1\\r\theta_2\end{array}\right)=O(\beta )\left(\begin{array}{c}\theta_\sigma\\ \theta_\rho\end{array}\right),\nonumber
\ee
where he matrix $O(\beta)$ is
\be
O(\beta )=\left(\begin{array}{cc} \cos \beta &\sin \beta \\ -\sin \beta &\cos \beta \end{array}\right).
\ee
The parameters $r$ and $\beta$ (denoted $\alpha$ in Ref.\,\onlinecite{AG}) depend on the asymmetry and interactions between the two channels and are obtained from 
\be
\matn{W^2_1=u_1K_1U_\theta+u_2K_2^{-1}U_\varphi \\ W^2_2=u_2K_2U_\theta+u_1K_1^{-1}U_\varphi}
\ee
and
\be
r^2=\frac{W_1}{W_2},
\qquad
\tan 2\beta=\frac{2{W_1W_2}}{u_2^2-u_1^2}.
\ee
The Hamiltonian density becomes
\bea
\mathcal{H}&=&\frac{1}{2}\sum_{j=\sigma,\rho}u_j\Big[K\dn_j(\partial_x\varphi_j)^2+K_j^{-1}(\partial_x\theta_j)^2\Big]\\
&&+
\eta{Z}\cos 2\sqrt{\pi}\Big[(r^{-1}\cos\beta+r\sin\beta)\varphi_\sigma\\
&&\hspace{2cm}+(r^{-1}\sin\beta -r\cos\beta)\varphi_\rho\Big]\label{eq59}
\eea
$K_\sigma$, $K_\rho$ are the Luttinger parameters for the bosons $\theta_\sigma$ and $\theta_\rho$ and they are obtained from
\bea
u_\sigma K_\sigma=u_1K_1r^{-2}\cos^2\beta+u_2K_2r^2\sin^2\beta-U_\varphi\sin 2\beta\nonumber\\
u_\rho K_\rho=u_1K_1r^{-2}\sin^2\beta+u_2K_2r^2\cos^2\beta+U_\varphi\sin 2\beta\nonumber
\eea
and
\bea
u_\sigma K^{-1}_\sigma=u_1K^{-1}_1r^{2}\cos^2\beta+u_2K_2^{-1}r^{-2}\sin^2\beta-U_\theta\sin 2\beta\nonumber\\
u_\rho K^{-1}_\rho=u_1K^{-1}_1r^{2}\sin^2\beta+u_2K^{-1}_2r^{-2}\cos^2\beta+U_\theta\sin 2\beta\nonumber.
\eea
\subsection{Double channel-flip scattering term}
The $Z$ term in Hamiltonian (\ref{eqbos}}) 
corresponds to correlated scattering of two electrons from one channel to the other, so that the momentum is conserved (Fig.\,\ref{fig:Zprocess}) and may in principle gap out one superposition of the $\varphi_\sigma$ and $\varphi_\rho$ bosons. The scaling dimension of this interaction is
\bea
d_Z&=&\frac{(r^{-1}\cos\beta+r\sin\beta)^2}{K_\sigma}+\frac{(r^{-1}\sin\beta-r\cos\beta)^2}{K_\rho}\nonumber
\eea
We have calculated this dimension numerically and confirmed that $d_Z>2$ away from the bottom of the bands for a large class of interaction potentials. For the special case of $B\approx 0$ on the normal side that we are interested in, the two Fermi velocities are equal $v_{F1}=v_{F2}=v_F$ and we obtain $r=1$, $\beta=\pi/4$ and thus $d_Z=2/K_\sigma$. On the other hand, for $K_\sigma$ we obtain 
\be
K_\sigma=\frac{\pi v_F-{\tilde V(k_{F1}+k_{F2})}}{\pi v_F+{\tilde V(k_{F1}+k_{F2})}}<1.
\ee
Thus $d_Z>2$ and the bulk interchannel pair tunneling term is irrelevant and can be ignored at low energies. We expect this result to be still valid at small $B$.
\subsection{Conductance at the $A\times N$ and $N\times A$ fixed points}
As discussed in Ref.\,\onlinecite{AG}, the stable fixed points in the interacting case are $A\times N$ and $N\times A$, 
corresponding to perfect Andreev reflection in channel 1 and perfect normal reflection in channel 2 or vice versa. For 
the case of decoupled channels, the corresponding conductance at the $A\times N$ fixed point, for an infinite interacting wire \cite{Fidkowski12} is
$G=(2e^2/h)K_1$ where $K_1$ is the Luttinger parameter for channel 1, a measure of the bulk interaction strength, 
with $K_1<1$ for repulsive interactions. This can be readily generalized to the general case of coupled channels, analyzed in Ref.\,\onlinecite{AG}. In general, 
the conductance at the $A\times N$ fixed point is proportional to the Green's function $\braket{\theta_1(\tau ,0)\theta_1(0,0)}$, for 
free bosons evaluated 
with boundary conditions $\phi_1(0)=\hbox{constant}$, $\theta_2(0)=\hbox{constant}$. It follows 
from Eq.\,(D.18) of Ref.\,\onlinecite{AG} that 
\be G_{A\times N}=\frac{2e^2}{h}{r^{2}K_\sigma K_\rho \over \cos^2\beta K_\rho+\sin^2\beta K_\sigma }.\ee
Similarly
\be G_{N\times A}=\frac{2e^2}{h}{r^{-2}K_\rho K_\sigma \over \cos^2\beta K_\sigma+\sin^2\beta K_\rho }.\ee
For the case of decoupled channels,\cite{AG} $\beta =0$, $r^2K_\sigma =K_1$, $r^{-2}K_\rho =K_2$,  giving the expected results. 
It can be seen \cite{AG} that $e^2/(hG_{A\times N})$ and $e^2/(hG_{N\times A})$ are the RG scaling dimensions of the tunnelling 
terms to the MM, $t_1$ and $t_2$ respectively. It then follows that, in the $\epsilon$-expansion limit, $G_{A\times N}=G_{N\times A}=e^2/h$.

\subsection{Conductance at QCP}
\begin{figure}[tp!]
\includegraphics[width=0.8\linewidth]{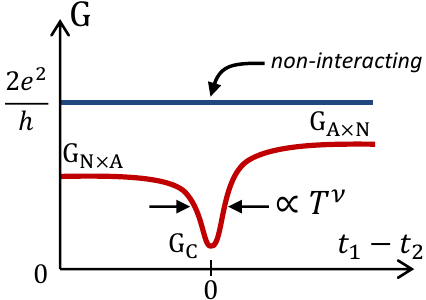}
\caption{\small\raggedright (color online) Schematic of the conductance as a function of the coupling asymmetry $t_1-t_2$. Whereas non-interacting theory (blue) predicts a coupling independent quantized conductance, the interacting theory (red) predicts a suppression of the conductance at the quantum critical point $t_1=t_2$.\label{fig:Gschem}}
\end{figure}

The attentive reader might be wondering how the critical point with equal tunnelling to both channels could be experimentally 
detected since  only the total conductance is readily measured. Ignoring interactions, this total conductance is completely 
independent of the relative tunnelling amplitude at zero temperature, having the universal value $2e^2/h$.~\cite{KA} Ref.\,\onlinecite{AG} showed 
that this changes radically when interactions are introduced. For an infinitely long normal wire and generic parameters, 
the zero temperature conductance is maximal to one channel and zero to the other. This corresponds to a simple type 
of conformally invariant boundary conditions in the bosonized theory
corresponding to perfect Andreev 
reflection boundary conditions for one channel and perfect normal reflection for the other, referred to as $A\times N$ and $N\times A$
boundary conditions.\,\cite{AG}
The conductances at the $A\times N$ and $N\times A$ fixed points, $G_{A\times N}$ and $G_{N\times A}$, depend on the 
bulk Luttinger parameters. As the relative strength of the tunnelling to the two channels is varied, 
a quantum critical point (QCP) occurs when they exactly balance. At this QCP the zero temperature conductance attains 
a special universal value, $G_C$, depending only on the bulk parameters of the 2-channel Luttinger liquid. So far $G_C$ has only been obtained [\onlinecite{AG}] using ``$\epsilon$-expansion'' methods 
when the repulsive interactions in the quantum wire are strong enough that the tunnelling to the MM is barely relevant 
in the renormalization group sense, with the tunnelling from channel $j$ having scaling dimension $\epsilon_j$ where $0<\epsilon_j \ll 1$.
 In this limit, $G_C=(e^2/h)(2\pi )^2(\epsilon_1+\epsilon_2)/{\cal F} (\nu )\ll e^2/h$. Here $\nu$ is one remaining parameter depending on 
 the bulk interactions in the LL, and the function $\cal F(\nu)$ was calculated in Ref.\,\onlinecite{AG}.
 On the other hand, in this limit, $G_{A\times N}=G_{N\times A}=e^2/h$ as we saw in the previous section. 
 For realistic interaction strengths 
 we thus might expect the zero temperature conductance to have two different constant values, depending on 
 which channel couples more strongly to the MM, with an abrupt dip at the QCP (Fig.\,\ref{fig:Gschem}). 
 Calculating the value of the conductance at the QCP accurately 
for general bulk parameters should be possible using the Density Matrix Renormalization Group technique.\,\cite{Rahmani} Alternatively, 
it is possible that the exact value could be found using boundary conformal field theory methods. More realistically, the 
normal part of the quantum wire will have a finite length, $L$, eventually contacting an Ohmic contact. This introduces another 
energy scale $\bar v/L$ where $\bar v$ is of order the velocities in the interacting wire. Ignoring back-scattering 
at this contact, below this energy scale, the conductance 
is expected to exhibit non-interacting behaviour, taking the universal value $2e^2/h$.  Observing the QCP still remains 
possible at energy scales, $E\gg \bar v/L$, where $E$ can be controlled by source-drain voltage, temperature or frequency.\,\cite{Fn3}

\subsection{Disorder and boundary interactions}
Finally, we turn to the combined effects of bulk interactions in the normal wires and disorder near the SN junction. Since we are discussing 
the low energy conductance, we may use the low energy effective Hamiltonian. Assuming the 
wire is clean beyond a length scale $\ell$, at energy scales $E\ll \bar v/\ell$,  we may represent the disorder by 
\be H_d=\psi^\dagger (0)\bb M\psi (0)+
[\Delta_b\psi_\uparrow (0)\psi_\downarrow (0)+h.c.].\label{Hd}\ee
In the non-interacting case, we have recently shown explicitly~\cite{KA} that $H_d$ has no effect 
on the zero energy conductance. Now consider the interacting case. 
The stability of the $A\times N$ and $N\times A$ fixed points 
against arbitrary boundary interactions, for a range of bulk LL parameters, was proven in Ref.\,\onlinecite{AG} where it was conjectured that whenever 
the $N\times N$ fixed point (corresponding to zero tunnelling to the MM) is unstable and both $A\times N$ and $N\times A$ 
fixed points are stable, the QCP exists.  This argument can be further substantiated by analyzing the RG scaling 
dimensions of perturbations at the QCP using the $\epsilon$-expansion. These are only perturbed by $O(\epsilon )$
from their values at the $N\times N$ fixed point.  As shown in Ref.\,\onlinecite{AG}, for small positive $\epsilon$ there is 
one further condition on the bulk LL parameters, $|\nu |<1$, for the QCP to occur at weak coupling, where
\be \nu \equiv {\sin 2\beta \over 2}\left({1\over K_\rho}-{1\over K_\sigma}\right).\label{nu}\ee
The $M_{12}$ interaction in Eq.\,\pref{Hd}, corresponding to inter-channel normal 
reflection, has RG scaling dimension $2+\nu$ and the $\Delta$ interaction, corresponding to inter-channel Andreev reflection, 
has dimension $2-\nu$, at small $\epsilon_i$.  Thus they are both irrelevant (with dimension $>1$) whenever the QCP occurs at weak coupling. 
This eliminates 4 of the 6 real parameters in $H_d$. The other two terms in $H_d$ are proportional 
to $M_{11}d\theta_1/dx+M_{22}d\theta_2/dx$ in the bosonized model. These can be eliminated from the Hamiltonian by shifting the $\theta_i(x)$ 
fields by step functions:
\be \tilde \theta_i(x)=\theta_i(x)+\zeta_iH (x)\ee
with $\zeta_i\propto M_{ii}$ and $H(x)$ the Heaviside step function. This shift can be used to eliminate 
these terms from the Hamiltonian {\it before} treating the coupling to the MM, so that they do not need to be considered at the QCP.

\section{Conclusion}
Confirming the existence of a Majorana mode at a topological SN junction remains an outstanding experimental challenge. 
We have shown how its presence might be confirmed by probing a novel critical point which it induces when the 
corresponding Majorana mode couples with equal strength to both channels of a Luttinger liquid and suggested an 
experimental set-up for doing so.

\acknowledgements 
We would like to thank  J.~Folk, D.~Giuliano and A.~Rahmani for helpful discussions. This research was supported in part by 
NSERC, CIfAR and the Swiss National Science Foundation. 
\appendix
\section{BdG equation and S-matrix}\label{sec:BdG}
\subsection{Bogliubov-de Gennes equations}
We introduce a 4-component spinor of fermion annihilation and creation operators
\be \Psi (x)\equiv \left(\begin{array}{c} \psi_\uparrow (x)\\ \psi_\downarrow (x) \\ 
\psi^\dagger_\uparrow (x) \\ 
\psi^\dagger_\downarrow (x) \end{array}\right).
\ee
These obey
\bea
\{\Psi_a (x) ,\Psi_b^\dagger (y)\}&=&\delta (x-y),\\  
\{\Psi_a (x) ,\Psi_b (y)\}&=&\tau^x_{ab}\delta (x-y),
\eea
where the indices $a$, $b=1,2,3,4$ and we introduce 4 component Pauli matrices, $\vec \tau$ which act on the particle-hole sectors
\be \tau^x\equiv \left(\begin{array}{cc} \bb 0 &\bb 1\\ \bb 1 & \bb 0\end{array}\right)\ee
et cetera. 
$\bb 0$ and $\bb 1$ are $2\times 2$ zero and unit matrices. In terms of these operators, the second quanitized 
Hamiltonian of Eq.\,(1) can be written
\be H={1\over 2}\int dx\Psi^\dagger(x) {\cal H}\Psi (x)+H_{int},\ee
where the Bogliubov-DeGennes (BdG) Hamiltonian is
\be {\cal H}=\left[ {p^2\over 2m}-\mu +V(x)+B(x)\sigma^z + \alpha p\sigma^y\right]\tau^z
+\Delta (x)\sigma^y\tau^y.\label{Hq6}\ee
Here $\sigma_i$ and $\tau_i$ for $i=x,y,z$ are pauli matrices in spin and particle-hole bases, respectively
and we have chosen $\Delta (x)$ real and positive for convenience, which can always be done by redefining the phases 
of the fermion fieids.  Note that the single-particle Hamiltonian ${\cal H}$ has the electron-hole symmetry 
\be
\tau^x{\cal H}\tau^x=-{\cal H}^*\label{Heh}
\ee
and also it is real ${\cal H}^*={\cal H}$, considering that $p=-i\hbar\partial_x$. The Rashba SOI corresponds to an electric field in the $z$-direction, between wire and the substrate. More generally it can point to other directions and as long as it is perpendicular to the magnetic field, the problem can be transformed to the convenient from given by Eq.\,\pref{Hq6} by a rotation around the magnetic field direction.

We first consider the eigenstates and eigenvalues of the BdG Hamiltonian, ${\cal H}$ in the asymptotic 
region, deep inside the normal wire where $\Delta (x)=V(x)=0$ and $B(x)=B$. Then the eigenstates and eigenvalues can be written 
in terms of those of the 2-component Hamiltonian describing electrons only (not holes)
\be {\cal H}_e\equiv {p^2\over 2m}-\mu +B\sigma^z + \alpha p\sigma^y.\ee
The energies are given by Eq.\,(2) of the paper and the corresponding 2-component wave-functions are $e^{ipx}\chi_{p,\pm}$, where up to normalization factors
\be \chi_{p,\pm}\propto \left(\begin{array}{c}i\alpha  p \\
B \mp \sqrt{(\alpha p)^2+B^2}
\end{array}\right).\label{cB}
\ee
We henceforth assume unit normalization
\be \chi_{p\pm}^\dagger \chi_{p\pm}=1.\ee
Note the important property
\be \chi_{-p\pm}=\chi^*_{p\pm}.\label{cp-}\ee
For a given energy, $E$, the 8 eigenstates of the BdG Hamiltonian in the large $x$ region are given by Eq.\,(2) of the paper 
and Eq.\,(\ref{cB}) with  momenta, $\pm p_{e\pm}$, $\pm p_{h\pm}$, for particle and hole solutions respectively. 
Here $p_{e/h\pm}$ are the positive momentum solutions of
\bea 
E&=&{p_{e\pm}^2\over 2m}\pm \sqrt{(\alpha p_{e\pm} )^2+B^2}-\mu \\
-E&=&
{p_{h\pm}^2\over 2m}\pm \sqrt{(\alpha p_{h\pm} )^2+B^2}-\mu .
\eea
The most general eigenstate of energy $E$ in the asymptotic region is ($p=\hbar k$)
\bea w(x)=\sum_{j=\pm} &\biggl[& a_{ej}\mat{\chi_{ej}\\0}
e^{-ik_{ej}x}
+a_{hj}\mat{0\\\chi_{hj}^{*}}e^{ik_{hj}x}\nonumber\\
&+&b_{ej}\mat{\chi_{ej}^{*}\\ 0}e^{ik_{ej}x}
+b_{hj}\mat{0\\ \chi_{hj}}e^{-ik_{hj}x}
\biggr].\label{Pas}\nonumber
\eea
Here $w$ is a 4-component spinor and $a^{e/h}$ and $b^{e/h}$ label left-moving and right-moving electron/hole quasi-particles, respectively. We used the short-hand notation $\chi_{ej}=\chi_{p_{ej},j}$. The conserved \emph{quasi-particle current} for the BdG Hamiltonian is
\bea 
J(x)&=&{-i\hbar \over 2m}\left[ w^\dagger(x) \tau^z{dw(x)\over dx}-{dw^\dagger(x)\over dx}\tau^z w(x)\right]\nonumber\\
&&\hspace{3.5cm}+\alpha w^\dagger(x) \tau^z\sigma^yw(x).\qquad\label{Jdef}
\eea
Note that $J(x)$ can be written in terms of matrix elements of the velocity operator
\be {dx\over dt}=v =-i[x,{\cal H}]=\left[ {p\over m}+\alpha \sigma^y\right]\tau^z.\ee
It can easily be proven that $dJ/dx=0$ for all $x$ for any wave-function, $w(x)$ which is an eigenstate 
of the BdG Hamiltonian, ${\cal H}w=Ew$. The current can be expressed in terms of the coefficients 
appearing in the large $x$ expression for $w(x)$, given in Eq.\,(\ref{Pas})
\be J=\sum_{j=\pm} \Big[ v_{ej}\Big(\abs{b_{ej}}^2-\abs{a_{ej}}^2\Big)+v_{hj} \Big(\abs{b_{hj}}^2-\abs{a_{hj}}^2\Big)\Big],\label{Jas}
\ee
where $v_{ej}\equiv v_j(p_{ej})$ and $v_{hj}\equiv v_j(p_{hj})$ for $j=\pm$ in which
\be v_{\pm} (p)={dE_\pm \over dp}=p\left[{1\over m}\pm {\alpha^2 \over \sqrt{(\alpha p)^2+B^2}}\right].\ee
To derive Eq.\,(\ref{Jas}) we have used
\be \chi^\dagger _{-p,\pm}\sigma^y\chi_{p,\pm}=0,
\ee
which is easily checked from Eq.\,(\ref{cB}) and
\be \chi_{p_+,+}^\dagger \left[{(p_++p_-)\over 2m}+\alpha \sigma^y\right]\chi_{p_-,-}=0\label{v+-}\ee
for any two momenta, of either sign, obeying
\be {p_+^2\over 2m}+ \sqrt{(\alpha p_+)^2+B^2}={p_-^2\over 2m}- \sqrt{(\alpha p_-)^2+B^2}
\ee
which is proven below.  Thus the 12 off-diagonal terms obtained by substituting the asymptotic expression  Eq.\,(\ref{Pas}) into 
the definition of the current, Eq.\,(\ref{Jdef}), all vanish, leaving only Eq.\,(\ref{Jas}). 
Since we consider $|E|\ll \Delta$, $w(x)$ decays exponentially to zero for $x\ll 0$ implying that the current is 
\be J(x)=0,\ \  (\hbox{for all}\ x).\label{J0}\ee
\emph{Proof} -
Here we give a proof of  Eq.\,(\ref{v+-}), which was used to evaluate the current. Defining
\be
\Gamma_{1,2}(p)\equiv B\pm\sqrt{(p\alpha)^2+B^2}
\ee
we can write
\bea 
&&\chi_{p_+,+}^\dagger \left[{(p_++p_-)\over 2m}+\alpha \sigma^y\right]\chi_{p_-,-}\propto {p_++p_-\over 2m}\times\nonumber \\
&&\Big[\alpha^2p_-p_++
\Gamma_1(p_-)\Gamma_2(p_+)
-\alpha^2p_-\Gamma_2(p_+)-\alpha p_+^2\Gamma_1(p_-)\Big].\nonumber
\eea
Note that $\chi_1$ and $\chi_2$ are {\it not} orthogonal spinors, being eigenstates of $(B\sigma^z+\alpha p\sigma^y)$ 
with {\it different} values of $p$.  
We use
\be {p_\pm^2\over 2m}=E\mp \sqrt{(p_+\alpha )^2+B^2}\label{p+-}\ee
to rewrite this as:
\bea 
&&\chi_{p_+,+}^\dagger \left[{(p_++p_-)\over 2m}+\alpha \sigma^y\right]\chi_{p_-,-}
\propto (p_++p_-)\times \nonumber \\
&&\hspace{2.5cm}\left[ \alpha^2(E-B)+{1\over 2m}\Gamma_1(p_-)\Gamma_2(p_+)\right]\qquad\label{v+-2}
\eea
Next, solving Eq.\,(\ref{p+-}), we find
\be {p_\pm^2\over 2m}=E+m\alpha^2\mp\sqrt{(E+m\alpha^2)^2+B^2-E^2}\ee
and
\be \sqrt{(p_\pm\alpha )^2+B^2}=\mp m\alpha^2+\sqrt{(E+m\alpha^2)^2+[B^2-E^2]}.\nonumber\ee
Substituting this into Eq.\,(\ref{v+-2}) gives zero.
\subsection{S-matrix}
The right-moving components of the asymptotic wave-function are linearly related to the left-moving components by the $4\times 4$ reflection 
matrix $\tilde {\bb r}$ defined by $\vec b=\tilde {\bb r}\vec a$.
Note that $\tilde {\bb r}$ can be decomposed into four $2\times 2$ blocks where $\tilde {\bb r}^{eh}$ gives the  amplitude for 
a right-moving hole to reflect as a left-moving electon et cetera:
\be \tilde{\bb r}=\left(\begin{array}{cc}
\tilde{\bb r}^{ee}& \tilde{\bb r}^{eh}\\
\tilde{\bb r}^{he}&\tilde{\bb r}^{hh}
\end{array}\right).\ee
Requring Eq.\,(\ref{J0}) to be true for arbitrary incoming wave-function amplitudes, $\vec{a}$ implies the conditions on the 
reflection matrix:
\be \sum_j\tilde{\bb r}^\dagger_{ij}v_j\tilde{\bb r}_{jk}=\delta_{ik}v_i\ee
where we have defined
\be \left(\begin{array}{c}
v_1\\v_2\\v_3\\v_4\end{array}\right)\equiv \left(\begin{array}{c}
v_+(p_{e+})\\v_-(p_{e-})\\v_+(p_{h+})\\v_-(p_{h-})\end{array}\right).\label{run}
\ee
It is convenient to define a unitary rescaled reflection matrix:
\be  {\bb r}_{ij}\equiv \sqrt{v_i\over v_j}\tilde{\bb r}_{ij}.\label{rtildef}\ee

The electron-hole symmetry property of the BdG Hamiltonian, imply that the BdG wave-functions of positive and negative energies
are related by:
\be w_{-E}(x)=\tau_x w_E^*(x).\ee
Eq.\,(\ref{cp-}) then implies that the amplitudes of the asymptotic wave-function in Eq.\,(\ref{Pas}) are related by
\be
\vec a(-E)=\tau^x\vec a^*(E),
\qquad 
\vec b(-E)=\tau^x\vec b^*(E).
\ee
Thus the reflection matrix obeys the electron-hole symmetry:
\be \tilde{\bb r}(-E)=\tau^x\tilde{\bb r}^*(E)\tau^x\ee
or equivalently:
\be \tilde{\bb r}^{ee}(E)=\tilde{\bb r}^{*hh}(-E),\ \  \tilde{\bb r}^{eh}(E)=\tilde{\bb r}^{*he}(-E).\ee
Noting that $v_{e\pm}(E)=v_{h\pm}(-E)$, we see that the same relation is obeyed by $ {\bb r}$:
\be  {\bb r}(-E)=\tau^x {\bb r}^*(E)\tau^x\ee
\subsection{Open boundary condition}\label{sec:obc}
When the normal wire is disconnected from the superconductor $\bb{r}$ is block-diagonal and $\bb{r}^{ee}$ is given by
\bea
\tilde r_{11}&=&
\frac{k_{F1}(\eps_2+B-E)+k_{F2}(\eps_1+B-E)}{k_{F1}(\eps_2+B-E)-k_{F2}(\eps_1+B-E)}\nonumber\\
\tilde r_{21}&=&\frac{-2k_{F1}(\eps_1+B-E)\sqrt{\frac{\alpha^2k_{F2}^2+(\eps_2+B-E)^2}{\alpha^2k_{F1}^2+(\eps_1+B-E)^2}}}{k_{F1}(\eps_2+B-E)-k_{F2}(\eps_1+B-E)}
\eea
The other two components can be obtained from $1\lr 2$ substitution. The wavefunction of the two channels in the normal side, with open boundary conditions are
\bea
f_{1E}(x)\propto\chi\dn_{1k_{F1}}e^{-ik_{F1}x}+r_{11}\chi^*_{1k_{F1}}e^{ik_{F1}x}+r_{21}\chi^*_{2k_{F2}}e^{ik_{F2}x}\nonumber\\
f_{2E}(x)\propto\chi\dn_{2k_{F2}}e^{-ik_{F2}x}+r_{22}\chi^*_{2k_{F2}}e^{ik_{F2}x}+r_{12}\chi^*_{1k_{F1}}e^{ik_{F1}x}\nonumber
\eea
One special case is when $B=0$, for which $r_{ii}=0, r_{12}=-1$ and $\chi_{1}=\chi_{2}^*$ independent of the momentum. Hence
\bea
f_{1E}(x)&=&\frac{1}{2}\mat{1\\ i}\Big[e^{-ik_{F1}x}-e^{ik_{F2}x}\Big],\nonumber\\
f_{2E}(x)&=&\frac{1}{2}\mat{1\\ -i}\Big[e^{-ik_{F2}x}-e^{ik_{F1}x}\Big],\nonumber
\eea
and they obey $f_{1E}^*(x)=-f_{2E}(x)$.
\section{Numerical analysis}\label{sec:num}

Our starting point is to represent the continuous model in Eq.\,1 by the following tight-binding model
\be
H=\frac{1}{2}\sum_n\Big[C_n\dg\mathbb{h}_nC_n+(C\dg_n\mathbb{t}_{n,n+1}{C}_{n+1}+h.c.)\Big]\label{eqtbmodel}
\ee
in which $C\dg_n=\left(\bmx{cccc} c\dg_{n\ua} & c\dg_{n\da} & c\dn_{n\ua} & c\dn_{n\da}\emx\right)$ and
\bea
\qquad
\bb{h}_n&=&(-\mu+g\mu_B B_n\sigma_z/2)\tau_z-\Delta_n\tau_y\sigma_y\label{eqmodel2a}\\
\bb{t}_n&=&(-t-i\alpha\sigma_y/2)\tau_z
\label{eqmodel2b}
\eea
Denoting BdG quasi-particles by $\psi\dg_\alpha=\sum_jC\dg_jw_{j\alpha}$ with energy $E_{\alpha}$, we can diagonalize the particle-hole symmetric Hamiltonian if $w_{j\alpha}$ satisfies the BdG equation
\be
E_{\alpha}w_{j\alpha}=\mathbb{h}_jw_{j\alpha}+\mathbb{t}_{j,j+1}w_{j+1,\alpha}+\mathbb{t}\dg_{j-1,j}w_{j-1,\alpha}
\ee
In the translationally-invariant case, $k$ is a good quantum number ($w_{nk}=w_ke^{ink}$) and the dispersion is obtained by the solution to $\mathcal{H}_kw_{k}=E_{k}w_k$ where
\be
\mathcal{H}_k=\bb{h}+\bb{t}e^{ik}+\bb{t}\dg e^{-ik}.
\ee
\subsection{Self-energies}
Green's function matrices can be defined for the $C_n$ operators using $\bb{G}_{nm}(t)\propto\braket{C\dn_n(t)C\dg_m(0)}$. In particular the local density of states can be extracted from the diagonal elements of the retarded function. For a semi-infinite chain defined on sites $n\ge1$ with site-independent parameters $\bb{h}$ and $\bb{t}$, this Green's function obeys the relation
\be
{\mathbb G}^R_{11}(\omega)=[(\omega+i\eta)\mathbb{1}-\mathbb{h}-\mathbb{\Sigma}_{11}^R(\omega)]^{-1}. \label{eqdos}
\ee
The self-energy is produced by electrons tunnelling to the second site, spending some time there or possibly moving to other sites and back. Removing the first site, we get the same chain we began with and therefore, the the self-energy is 
\be \mathbb{\Sigma}^R_{11}(\omega)=\mathbb{t}\mathbb{G}^R_{11}(\omega)\mathbb{t}\dg, \label{selfcon}\ee
producing a recursive equation for the Green's function matrix of the first site. 

Eq. (\ref{selfcon}) can be conveniently derived, for example, from an imaginary time Feynman path integral representation.  Since the action is quadratic, 
we may exactly integrate out the fermion fields on all sites $n\geq 2$.  This produces an extra term in the effective action for site $1$:
\be \delta S_{{\rm eff} 1}=\sum_nC_1^\dagger (i\omega_n)\bb{t}{\mathbb G}_{22}^{M'}(i\omega_n)\bb{t}^\dagger C_1(i\omega_n)\ee
where ${\mathbb G}^{M'}_{22}(i\omega_n)$ is the Matsubara Green's function for the fermions at site $2$ in a semi-infinite chain {\it  beginning at site } $2$ 
and $\omega_n\equiv \pi (2n+1)T$. 
However, since the chain is semi-infinite and $\bb{h}$ and $\bb{t}$ are site independent, it follows that
\be {\mathbb G}^{M'}_{22}(i\omega_n)={\mathbb G}^{M}_{11}(i\omega_n).\ee
Thus we see that the self-energy of the Matusbara Green's function at site $1$ is
\be  \mathbb{\Sigma}^M_{11}(i\omega)=\mathbb{t}\mathbb{G}^M_{11}(\omega)\mathbb{t}\dg ³ .\ee
Continuing the Matsurbara Green's function to real frequencies, $\omega$, gives the retarded Green's function, and hence Eq. (\ref{selfcon}). 

The solution to this recursive equation is obtained by finding eigenvalues/vectors of the matrix $\bf{D}$ defined as
\be
\mathbf{D}\equiv\mat{\mathbb{B} & -\mathbb{C} \\ \mathbb{1} &\mathbb{0}}, \qquad \mathbf{D}\vec u_i=\lambda_i\vec u_i, 
\ee
in which
\be
\mathbb{B}=\mathbb{t}\dg\mathbb{t}^{-1}(\omega\mathbb{1}-\mathbb{h}_0)(\mathbb{t}\dg)^{-1}, \qquad
\mathbb{C}=\mathbb{t}\dg\mathbb{t}^{-1}
\ee
These eigenvalues/vectors are arranged in 
matrices $\mathbf{U}$ and $\mathbf{\Lambda}$ according to
\be
\bf{U}=\mat{\vec u_i}=\left(\bmx{c|c}\mathbb{U}_{11}&\mathbb{U}_{12}\\\hline \mathbb{U}_{21} & \mathbb{U}_{22}\emx\right), \quad \mathbf{\Lambda}=\mat{\lambda_i}=\left(\bmx{c|c}\mathbb{\Lambda}_1&\mathbb{0}\\\hline \mathbb{0} & \mathbb{\Lambda}_2\emx\right)\nonumber
\ee
so that $\mathbb{\Lambda}_1<1$. 
Then the solutions is
\be
\mathbb{G}=(\mathbb{t}\dg)^{-1}\mathbb{U}_{11}\mathbb{\Lambda}_1\mathbb{U}_{11}^{-1},\label{eq120} 
\ee
{\it Proof} - First, note that the recursive formula for the Green's function implies that $\bb{t}\dg\bb{G}$ satisfies the equation
\be
(\bb{t}\dg\bb{G})^2-\mathbb{B}(\bb{t}\dg\bb{G})+\mathbb{C}=\mathbb{0}.\label{eq121}
\ee
Writing the two components of the Eq. $\mathbf{D}\mathbf{U}=\mathbf{U}\mathbf{\Lambda}$,
\be
\mathbb{B}\mathbb{U}_{11}-\mathbb{C}\mathbb{U}_{21}=\mathbb{U}_{11}\mathbb{\Lambda}_1
\andd \mathbb{U}_{11}=\mathbb{U}_{21}\mathbb{\Lambda}_1
\ee
Combination of these gives
\be
\mathbb{B}\mathbb{U}_{21}\mathbb{\Lambda}_1\mathbb{U}_{21}^{-1}-\mathbb{C}\mathbb{U}_{21}\mathbb{U}_{21}^{-1}=\mathbb{U}_{21}\mathbb{\Lambda}_1\mathbb{\Lambda}_1\mathbb{U}_{21}^{-1} 
\ee
which can be written as of Eq.\,\pref{eq121}
\be
\Big(\mathbb{U}_{21}\mathbb{\Lambda}_1\mathbb{U}_{21}^{-1}\Big)^2-\mathbb{B}\Big(\mathbb{U}_{21}\mathbb{\Lambda}_1\mathbb{U}_{21}^{-1}\Big)+\mathbb{C}=\mathbb{0}.
\ee
Comparing the two and noting that $\mathbb{U}_{21}\mathbb{\Lambda}_1\mathbb{U}_{21}^{-1}=\mathbb{U}_{11}\mathbb{\Lambda}_1\mathbb{U}_{11}^{-1}$ we arrive at the result of Eq.\,\pref{eq120}.

\begin{figure}
\centering
\includegraphics[scale=0.8]{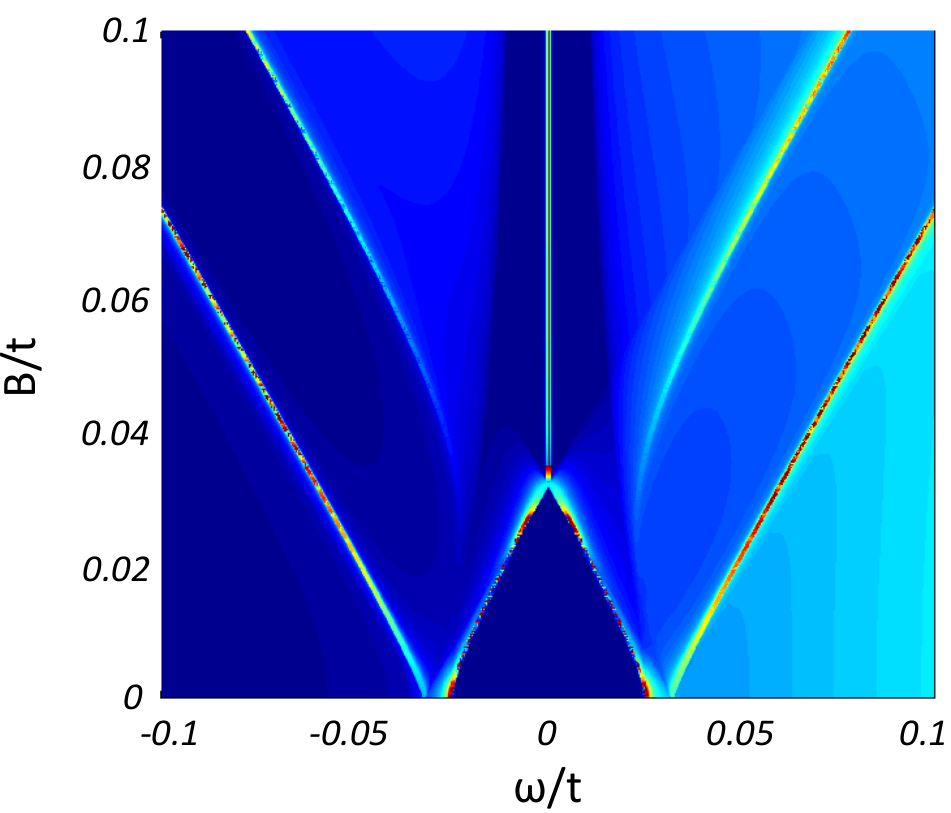}
\caption{
\raggedright\small (color online) Local density of states $\varrho_{1}(\omega)$ (of electrons with both spins) at the end of a semi-infinite chain as a function of energy $\omega/t$ and Zeeman energy $B/t$, obtained as an algebraic solution to Eq.\,\pref{eqdos}. The dark (bright) area corresponds to low (hight) local density of states. The model is that of Eq.\,(\ref{eqtbmodel}-\ref{eqmodel2b}) with parameters $\alpha/t=0.14$, $\Delta/t=0.025$ and $\mu/t=0.02$. The s-wave gap at zero $B$ closes by the magentic field and re-opens as a p-wave gap supporting a Majorana bound state at $\omega=0$.\label{fig:Ldos}
}
\end{figure}

Figure\,\ref{fig:Ldos} shows this function at the end of a semi-infinite chain with Rashba SOI and s-wave pairing as a function of energy and the Zeeman splitting, for the parameters considered in this paper. Note that the superconducting gap closes by turning on the Zeeman field and re-opens with an additional MM at zero energy.
\subsection{Exact diagonalization and the reflection matrix}
We consider an SN junction embedded in an infinite system. The effect of normal/superconductor semi-infinite leads can be taken into account using the self-energy matrix. This method is an effective way to integrate out both gaped and gapless leads and it helps to avoid the problem of finite size effect and/or the necessity of diagonalizing very large matrices. The solution to
\be
[(\omega+i\eta)\mathbb{1}-\mathbb{H}-\mathbb{\Sigma}^R(\omega)]w=s
\ee
for $s=0$ gives eignfunctions of the Hamiltonian at all sites inside the system. For scattering problems, we are interested to know the reflection amplitudes as a result of some incident waves $w=w_{ref}+w_{in}$. It can be shown that including the source term
\be
s=-[\mathbb{\Sigma}^R(\omega)-\mathbb{\Sigma}^A(\omega)]w_{in}
\ee
does the job. Having obtained $w_{ref}$, the reflection amplitudes can be read off by multiplying the reflected wave by the eigenfunctions. Conductance can then be calculated from the BTK formula. 
\subsection{Parameters}\label{sec:param}
For the parameters we use similar values as those reported in Ref.\,\onlinecite{Mourik12}. In this experiment an InSb nanowire has been used with the effective mass $m^*=0.015m_e$. The proximity to NbTiN superconducting electrodes produced an induced gap of $\Delta\sim 250\mu$eV. The $g$-factor is $g=50$ and $B_S\sim 0.15{\rm T}\sim 217\mu$eV (we take $g\mu_B/2=1$). Also $\alpha\sim 0.21$eV\AA, leads to $E_{SO}=m^*\alpha^2/2\hbar^2\sim 50\mu$eV. In order to simulate this system with a tight-binding model, we note that a tight-binding model with hopping parameter $t$ and lattice constant $a$, at low filling fractions approximates a parabolic dispersion with the effective mass $m^*=\hbar^2/2ta^2$. For example taking $t=10$meV and $a=15$nm produces the desired effective mass.\,\cite{Rainis13} Therefore, measuring all the energy scales and length scales in units of $t$ and $a$, respectively, we set $\alpha/ta=0.14$, $\Delta/t=0.025$, $\mu/t=0.02$. However, for the B-field we choose $B/t=0.05$ in the superconducting side to be deep into the topological state as opposed to the too low value of 0.0217 used in the experiment \cite{Mourik12} (Fig.\,\ref{fig:Ldos}).\\

\section{Additional experimental considerations}
In order to observe the predicted QCP, the temperature needs to be $k_BT\ll\hbar\Gamma$, where $\hbar\Gamma$ is the width of the zero bias peak, set by the strength of the coupling $\Gamma\sim\sqrt{t_1^2+t_2^2}$. Note that the disorder in the system does not affect the zero temperature conductance but it strongly modifies $\Gamma$ which sets the height of the ZBP at $k_BT\sim \hbar\Gamma$ and therefore, nanowire has to be sufficiently clean. A good non-interacting check would be to also set $k_BT\ll \hbar v_F/L$,  where $L$ is the length of the nanowire, and recover the non-interacting linear conductance $2e^2/h$. This quantized conductance has not been observed so far in topological SN junction experiments on InSb\,\cite{Mourik12,Deng12} or InAs.\cite{Das12} Based on our studies,~\cite{KA} we expect the suppression of the conductance from the quantized value to be caused by either too large temperature (compared to the coupling broadening).

Observing quantized conductance is a prior to an unambiguous detection of the QCP. Nevertheless, once a quantized ZBP is measred, QCP can be used to role out the other possible origins of the ZBP.
In order to see the effect of interactions and the QCP, either the bias or the temperature has to be increased $max(V,T)>\hbar v_F/L$. The suppression of the conductance from $2e^2/h$ due to interaction is set by the Luttinger parameters of the nanowire. These are set by the ratio of the screened Coulomb repulsion by the Fermi velocity. The latter depends on the value of the density and the effective mass of the nanowire. 
As in 1D $k_F=n\pi/4$, we have $v_F=h n/8m^*$. For InSb nanowires with the density $n\sim 8\times 10^8$ $m^{-1}$,\,\cite{Plissard} we get $v_F\sim 5\times 10^6$ $m/s$. For a nanowire of length $L\sim 1\mu m$, this gives a mean level spacing of $\hbar v_F/L\sim 20 K$, below which non-interacting results are dominant. The observation of QCP at low temperatures thus requires lower densities or materials with larger effective mass. A promising system could be cleaved edge overgrown quantum wires~\cite{Quay} which provide clean ballistic wires with a controllable number of channels and strong spin-orbit interaction.



\begin{thebibliography}{50}
\bibitem{sob} For simplicity we focus on the case of orthogonal magnetic field and spin orbit vector, $\vec \alpha$ in 
this paper. However, we expect similar behavior in the more general case where they are not orthogonal. 
(Y. Komijani and I. Affleck, in progress.)
\bibitem{Fn1}
In addition to $\mu>B$, the case of $-(m^2\alpha^4+B^2)/2m\alpha^2<\mu<-B$ also offers a small window of two bands crossing the zero energy, which we neglect here.
\bibitem{Fn2}
For a single channel this has a trivial extension. The conductance becomes $G=e^2/h[1-\det(\bb r_M)]$ which is either $0$ or $2e^2/h$ and robust against disorder near the junction.
\bibitem{Fn3}
At temperatures larger than the zero bias peak width $\Gamma\sim\sqrt{t_1^2+t_2^2}$ the height of $dI/dV$ is suppressed. Thus, we assume $T\ll\Gamma$ in this paper.
\end{thebibliography}
\end{document}